\documentstyle[aps,prl,epsf,floats]{revtex}

\newcommand{\be}{\begin{equation}}
\newcommand{\ee}{\end{equation}}
\newcommand{\bea}{\begin{eqnarray}}
\newcommand{\eea}{\end{eqnarray}}

\begin{document}

\ifpreprintsty \else
\twocolumn[\hsize\textwidth\columnwidth\hsize\csname@twocolumnfalse%
\endcsname \fi

\draft

\title{Fresnel laws at curved dielectric interfaces of microresonators}

\author{M. Hentschel and H. Schomerus}
\address{
Max-Planck-Institut f\"{u}r Physik komplexer Systeme,
N\"{o}thnitzer Str. 38, 01187 Dresden, Germany
}

\date{\today}
\maketitle

\begin{abstract}
We discuss curvature corrections to Fresnel's laws for the
reflection
and transmission of light at a non-planar refractive-index boundary.
The reflection coefficients are obtained from the resonances of a
dielectric disk within a sequential-reflection model.
The Goos-H\"anchen effect for curved light fronts at a planar
interface can be adapted to provide a qualitative and
quantitative extension of the ray model which explains the observed
deviations from Fresnel's laws.

\end{abstract}

\ifpreprintsty \else
] \fi              %of \twocolumnfalse

%\section{Introduction}

The fabrication of lasing microresonators \cite{jensnature,gmachlscience}
and opto-mechanical microdevices \cite{ashkin}
has generated a surge of interest
in the confinement and propagation of light in small dielectric structures.
Some 
understanding has been achieved from the ray optics of these systems,
complemented by Fresnel's laws of refraction and reflection at
the interfaces, e.g., in order to identify and describe the relevant resonator modes
\cite{jensnature,gmachlscience}.
Fresnel's laws give the probability of
reflection and refraction of 
plane electromagnetic waves 
at planar interfaces of media with different refractive index $n$.
Microresonators, however, often are so small that the curvature of their boundary
cannot be neglected.

In this paper we investigate, in the limit of large wavenumbers,
the corrections 
to the Fresnel coefficients which appear due to the curvature of the
dielectric interface.
The reflection coefficients are obtained
via a sequential-reflection model \cite{procamst} from the resonance widths of
a microresonator, which are analytically accessible
for large wavenumbers. 
The deviations from Fresnel's laws
are most noticeable around the critical angle for total
internal reflection, $\chi_c = \arcsin (1/n)$ (where
the refractive index of the surrounding medium is set to unity) and
amount in a systematic reduction of the reflection probability.

The reduction of the reflection probability is conventionally related to
tunneling escape at the curved interface. In view of the
previous success of the
ray model, which often is desired to be retained
for its simplicity,
we provide an alternative  qualitative and quantitative  explanation of the deviations
by incorporating into this model the Goos-H\"anchen effect
\cite{ghs_entdeck,ghs_artmann,ghs_lai,snyder,ghs_horowitz,ghs_lotsch}.
This
effect
results from the interference of
rays in a beam with slight variations of the angle of incidence
and consists of 
a shift of the effective plane of reflection.
At a curved interface,
it turns out that the reflection probability is then reduced
because the angle of incidence at the effective interface is smaller than 
at the physical interface.

There is evidence obtained in the context
of quantum-mechanical scattering problems \cite{herb}
that incorporating the  Goos-H\"anchen effect
is equivalent to a semiclassical approximation.
Also the effect has been used in Ref.\  \cite{chowdhurychang}
to explain the decreased
spacing of  
resonances observed in experiments with
dielectric spheres, in terms of an
effective optical size of the cavity which is larger
than its physical size.
Our work can be seen as complementary to this previous study, because
we are concerned with the resonance widths, not only the resonance energies.

Curvature corrections to Fresnel coefficients
have been addressed in the past in a number of
works, by applying various techniques; see for example
Refs.\ \cite{fresnel_snyder,fresnel_derrest,fresnel_ruan,%
fresnel_derrest2}.
The works closest in spirit to the present paper are those
which employ the complex ray method, e.g.,
to describe light rays approaching
a disk from outside \cite{fresnel_ruan,fresnel_derrest2}.

Although we restrict our discussion to circular interfaces,
the results for the reflection coefficients should be applicable to microresonators
of general shape as soon as locally the curvature can be approximated as
a constant.

%%%%%%%%%%%%%%%%%%%%%%%%%%%%%%%%%%%%%%%%%%%%%%%%%%%%%%%%%%%%%%%%%%%%%%%%%%%
%\section{Exact reflection coefficients for curved interfaces}

According to Fresnel's laws, a plane 
electromagnetic wave incident on a planar dielectric interface  
with angle of incidence $\chi$ is reflected
with the polarization-dependent coefficients
\bea
R^{\rm TM} =  \frac{\sin^2(\chi-\eta)}{\sin^2(\chi+\eta)}\:,\quad
R^{\rm TE}  =  \frac{\tan^2(\chi-\eta)}{\tan^2(\chi+\eta)}
\label{freste} \:,
\eea
where
TM (TE) signifies transverse polarization of the magnetic (electric)
field at the interface and
$\eta = \arcsin (n \sin \chi)$ is the direction of the refracted beam
(according to Snell's law). 

Let us compare the Fresnel coefficients with the reflection coefficients
at a curved interface with radius of curvature $r_c$.
Their angular dependence can be conveniently  obtained 
from the 
energies and widths
of resonance states in
a two-dimensional circular disk 
of radius $r_c$.
In this geometry the two possible polarization directions decouple
and angular momentum (quantum number $m$) is conserved.
We introduce polar coordinates $r$ and $\phi$
and denote the (complex) wavenumber by $k$. We will concentrate on the
case close to geometric optics ${\rm Re}\,k r_c\gg 1$.

The resonance states are obtained by matching the 
wave field $\propto J_m(nkr)  e^{i m \phi}$ inside the disk (with the Bessel function $J$)
at $r=r_c$ 
to the wave field $\propto H_m^{(1)}(kr) e^{i m \phi}$ outside the disk
(with the Hankel function $H^{(1)}$),
where the matching conditions follow from Maxwell's equations:
\be \label{tmstartderiv}
J_m(nkr_c) \, {H_m^{(1)}}'(kr_c) = n \, J'_m (nkr_c) \, H_m^{(1)}(kr_c)\:,
\ee
for TM polarization, and
\be
\label{testartderiv}
n J_m(nkr_c) {H_m^{(1)}}'(kr_c) = J_m'(nkr_c) H_m^{(1)}(kr_c) \:,
\ee
in the TE case (primes denote derivatives).
Given a complex solution  $k$, the angle of incidence is
obtained
from the real part by comparing the
angular momentum in the ray picture and in the wave picture, 
\be
\label{sinchi}
\sin\chi =  \frac{m}{n \,{\rm Re}\,k r_c},
\ee
while  the reflection probability
\be
R  = \exp (4 n \,{\rm Im}\,k r_c \cos \chi )
\label{rcoeff}
\ee
follows from the imaginary
part of $k$ because it determines the escape rate, which in turn can be related to
$R$ by the sequential-reflection model \cite{procamst}.

Because the discrete set of resonance energies obtained
from
Eqs.\ (\ref{tmstartderiv}) and (\ref{testartderiv}) only is meaningful
for the disk, let us first derive 
analytical expressions for the resonance width
as a function
of a continuous resonance energy which
smoothly
interpolate between these solutions.
It is interesting to note \cite{procamst} that one cannot
simply expand Eqs.~(\ref{tmstartderiv}), (\ref{testartderiv}) in ${\rm
Im}\, k$ when $k$ is not close to an exact solution.
Moreover, for TE polarization ${\rm Im}\,k$ would diverge at the Brewster
angle if it is calculated by inserting the Fresnel coefficient (\ref{freste})
into Eq.\ (\ref{rcoeff}).
In order to achieve a more accurate expansion
we separate out the problematic term
and cast Eqs.~(\ref{tmstartderiv}) and (\ref{testartderiv}) into the form
\be\label{ratbess}
\frac{J_{m}'(nkr_c)}{J_{m}(nkr_c)} ={\cal F}(kr_c), \ee
with 
\bea {\cal F}^{\rm TE}(x)
=n \frac{{H_{m}^{(1)}}'(x)}{H_{m}^{(1)}(x)},\quad
{\cal F}^{\rm TM}(x)&=&n^{-2} {\cal F}^{\rm TE}(x),
\eea
depending on the polarization.
In both cases, ${\cal F}(kr_c)$ is a
slowly varying complex function of its argument, and the argument can
be taken real because ${\rm Re}\,k\gg|{\rm Im}\,k|$.
The logarithmic derivative of  Bessel functions, however, 
is a rapidly fluctuating function, and its dependence on ${\rm Im}\,k$
has to be worked out carefully.
This can be achieved by approximation by tangents \cite{gradstein},
\bea\label{debyecos}
&&\frac{J_{m}'(nkr_c)}{J_{m}(nkr_c)} =-\tan\alpha\cos\chi
,
\\
&&\alpha=m\cot\chi+m\chi-m\frac{\pi}{2}-\frac{\pi}{4}+i n \,{\rm Im}\, k r_c\cos\chi,
\eea
where
$\chi$ is given as a function of ${\rm Re}\,k$
by Eq.\ (\ref{sinchi}). We expanded $\alpha$ linearly in ${\rm Im}\, k$ and neglected
terms of order $({\rm Re}\,k r_c)^{-1}$.
Equation (\ref{ratbess}) now can be solved for $\alpha$, without any further
approximation.
>From the imaginary part one deduces
\begin{equation}
\label{imk}
-{\rm Im}\, k r_c=\frac{1}{n\cos\chi}\,{\rm Im}\,\arctan
\frac{\cal F}{\cos\chi}.
\end{equation}
Although this can already be taken as the final result, we
further may insert 
the uniform approximation
\cite{gradstein}
\be
{\cal F}^{\rm TE} % ({\rm Re}\, k r_c)
=
i n\cos\eta\left[1+\frac{1}{\sin^2\eta}
\left(\frac{K_{2/3}(z)}{K_{1/3}(z)}-1\right)\right]
\begin{figure}[!t]%[htb]
  \epsfxsize=\columnwidth
   \centerline{\epsffile{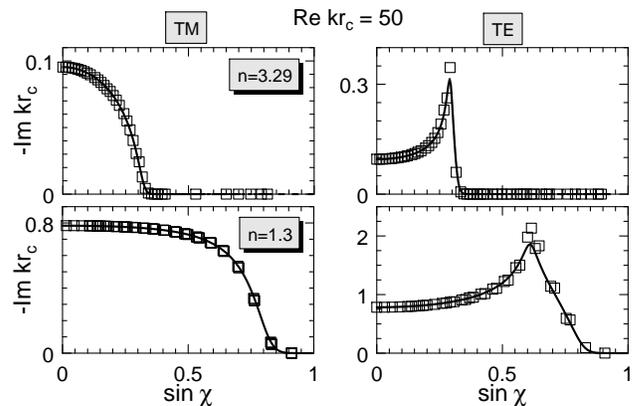}}
  \caption{Resonance widths $-\,{\rm Im}\,k r_c$ for a dielectric disk
	with ${\rm Re}\,
	kr_c=50$ and a refractive index $n=3.29$ (upper panel) and $n=1.3$
	(lower panel), for TM and TE polarized light, as a function of the
	angular momentum [parameterized according to
	Eq. (\ref{sinchi})].
	The analytical
  result from Eqs.\ (\ref{imk}), (\ref{unif})
  (solid curves) is compared with the
  exact results from Eqs.~(\ref{tmstartderiv}), (\ref{testartderiv}) (squares).
  }
  \label{fig_tantmte}
\end{figure}

,
\label{unif}
\ee
and similarly for ${\cal F}^{\rm TM}={\cal F}^{\rm TE}/n^2$,
with the modified Bessel function $K$, the angle of refraction $\eta$
(which is a complex number for $\chi>\chi_c$),
and $z=-i {\rm Re}\,k r_c\cos^3\eta/(3\sin^2\eta)$.
In Fig.~\ref{fig_tantmte} we illustrate that Eqs.\ (\ref{imk}), (\ref{unif})
agree very well with the exact solutions
of Eqs.\ (\ref{tmstartderiv}) and (\ref{testartderiv}),
even close to the Brewster angle for TE polarization,
and interpolates
smoothly in between.

\begin{figure}[!t]%[htb]
  \epsfxsize=7cm
  \centerline{\epsffile{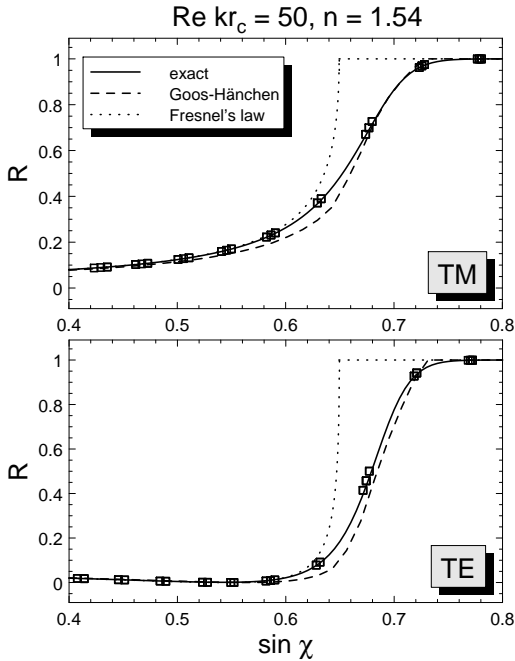}}
  \caption{Reflection coefficients $R$ in a dielectric
	disk with ${\rm Re}\,kr_c=50$. The solid curve is the
	analytical result
	(\ref{rres}), which smoothly interpolates between the exact solutions
  of Eqs.~(\ref{tmstartderiv}), (\ref{testartderiv}) with real part
  close to 50 [squares,  translated into angular-dependent reflection coefficients
	by Eqs.\ (\ref{sinchi}) and (\ref{rcoeff})].
	The dashed curve is the result of
	incorporating the Goos-H\"anchen effect into a ray model
	(assuming for TE polarization
	that the shift is the same as for TM polarization, 
	for reasons explained in the text).
	The dotted curve is
	Fresnel's law (\ref{freste}).
}
\label{figkr50}
\end{figure}

\begin{figure}[!t]%[htb]
  \epsfxsize=7cm
    \centerline{\epsffile{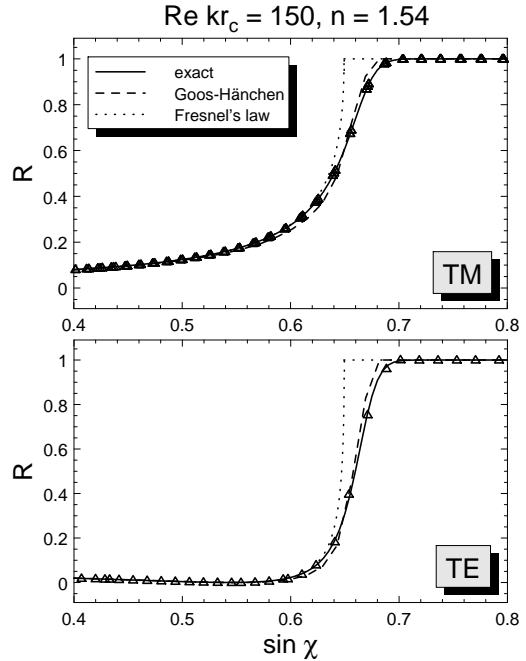}}
  \begin{center}
  \caption{Same as Fig.~\ref{figkr50} but for a wavenumber ${\rm Re}\,kr_c=150$.
	The  result based on the Goos-H\"anchen effect is almost
	indistinguishable from the exact result.
  }
  \label{figkr150}
  \end{center}
\end{figure}

The angular dependence
of the reflection coefficients can now be obtained by combining
Eqs.~(\ref{rcoeff}) and (\ref{imk}), giving the final result
\be
R=\left|
\frac{\cos\chi+i{\cal F}}{\cos\chi-i{\cal F}}
\right|^2
\label{rres}
.
\ee
Figures \ref{figkr50} and \ref{figkr150} show the result
for the two values
${\rm Re}\,kr_c=50$ and ${\rm Re}\,kr_c=150$, respectively.
Deviations from Fresnel's laws
(\ref{freste}) are most visible around the critical angle
$\chi_c = \arcsin (1 / n)$ where the reflection coefficients
increase rapidly as the regime of total internal
reflection is approached.
The correction consists not only
in a broadening of the transition interval, 
but most notably also in a shift of
this transition region towards higher angles of incidence, resulting
in a systematic reduction of the reflection coefficient.

%%%%%%%%%%%%%%%%%%%%%%%%%%%%%%%%%%%%%%%%%%%%%%%%%%%%%%%%%%%%%%%

%\section{Goos-H\"anchen effect}

The deviations from Fresnel's laws in Figs.~\ref{figkr50} and \ref{figkr150}
increase as ${\rm Re}\,kr_c$ is reduced,
that is, the more noticeable the curvature of the interface is.
On the other hand, 
in the zero-wavelength limit ${\rm Re}\,kr_c\to\infty$ of geometric optics
any interface appears planar, and Fresnel's laws 
should apply without modification.
Indeed it can be seen that they are
recovered from 
Eq.\ (\ref{rres})
when the approximation by tangents is also applied 
to the Hankel functions in ${\cal F}$,
resulting in ${\cal F}^{\rm TM} = in^{-1}\cos\eta$,
${\cal F}^{\rm TE} = i n\cos\eta$.
The deviations close to the critical angle are directly related
to the break-down of this approximation
when the argument of the Hankel functions
becomes smaller than the index.
As we will discuss now,
the curvature corrections to Fresnel's coefficients can be obtained within a
minimal extension of the ray picture when
the Goos-H\"anchen effect is taken into account (the result obtained
is given by the dashed curves in Figs.~\ref{figkr50} and \ref{figkr150}).

\begin{figure}[!t]%[htb]
  \epsfxsize=6cm
   \centerline{\epsffile{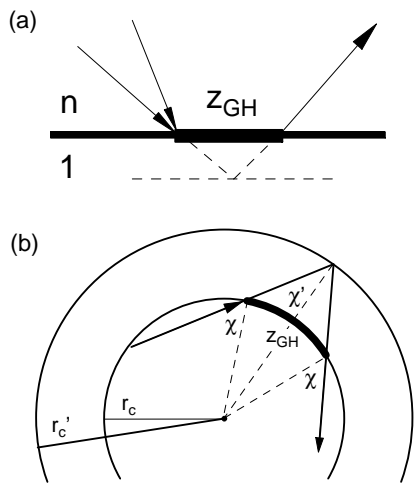}}
  \begin{center}
  \caption{(a) Goos-H\"anchen shift at a planar interface. An incident beam
  containing contributions from plane waves of
  slightly different angle of incidence $\chi$ appears to be
  reflected at a position that is shifted a distance $z_{\rm GH}$ away from
  the point of incidence. Alternatively, one can think of the beam to be
  reflected at a shifted interface indicated by the dashed lines.
	(b) Goos-H\"anchen effect at a curved interface.
		The reflection seems to occur at an interface of 
			larger curvature radius $r_c' > r_c$ under a smaller angle $\chi' < \chi$ of
				incidence.}
  \label{fig_ghsintro}
  \end{center}
\end{figure}

The Goos-H\"anchen effect
\cite{ghs_entdeck,ghs_artmann,ghs_lai,snyder,ghs_horowitz,ghs_lotsch} 
refers to the displacement 
of the reflected beam when the incident 
beam consists of rays with slight variations
of the angle of incidence, 
and arises because each ray experiences a slightly different phase
shift when it is reflected.
As is illustrated in Fig.~\ref{fig_ghsintro}a,
the lateral shift along the interface 
also can be interpreted as resulting from a
displacement of the effective plane of reflection
(a signature of this displacement is the
increased Wigner delay time which recently has been 
measured
at metallic gratings \cite{ghs_chauvat}.)

Since at a planar interface the reflection law
is not affected by parallel displacement and
fulfilled for the mean angles of incidence,
the reflection coefficients are not affected by the Goos-H\"anchen
effect---the only consequence of the slight variation of angles is that their
angular dependence is smeared out.
However, the situation
changes at a curved interface \cite{herb} as shown in Fig.~\ref{fig_ghsintro}b.
The intersection of the incident and the laterally shifted
reflected ray defines an
effective boundary of radius $r_c' > r_c$. 
We now can assume that the ray is specularly reflected at the effective
boundary, resulting
in a smaller effective angle of incidence $\chi' < \chi$, and
evaluate the Fresnel reflection coefficients
(\ref{freste}) at this smaller angle $\chi'$.
Since the reflection probability is then reduced
this qualitatively explains the observed
deviations from Fresnel's laws
in Figs.\ \ref{figkr50} and \ref{figkr150}.

For a quantitative comparison we need the
distribution of angles of incidence $P(\chi)$,
which can be related to the radial width
$\propto r_c(n{\rm Re}\,k r_c)^{-2/3}$
of the caustic in the radial wave field $J_m(n k r)$.
(The same distribution of angles will also be used to
smooth out the reflection coefficients
as is appropriate even for the planar interface.)
The variation of angles of incidence
arises from the intrinsic curvature of
the beam wave front and also directly from the curvature 
of the interface---note that both mechanisms should
contribute equally to the Goos-H\"anchen effect.
We found that 
for our purposes $P(\chi)$ is
sufficiently well approximated by 
that of a Gaussian beam of half width
$\sigma= r_c(n{\rm Re}\,k r_c)^{-2/3}\ll r_c$.
It is good to observe that this Gaussian beam geometry 
does not put us into conflict with the finite disk size
while at the same time leaving us in the paraxial regime
$n{\rm Re}\, k\sigma = (n{\rm Re}\, kr_c)^{1/3}\gg 1$.
For TM polarization the effective radius of curvature $r_c'$
can then be calculated by applying the formulas of Ref.\ \cite{ghs_lai}
(which are lengthy expressions and hence not given here).
The result of this Goos-H\"anchen-effect-based
approach is presented as
the dashed curves in the upper panels of Figs.~\ref{figkr50} and \ref{figkr150}.
There is good agreement with the exact results obtained within the
sequential-reflection model.
For TE polarization and the chosen refractive index,
the analytic results in the literature becomes applicable only for
${\rm Re}\, k r_c \gtrsim 1000$.
Surprisingly (see however Ref.\ \cite{haibel}), in the current situation
nice agreement is found by simply assuming that the Goos-H\"anchen shift
is identical to the TM case, 
as is indicated by the dashed curves in the lower panels
of  Figs.~\ref{figkr50} and \ref{figkr150}.
(For ${\rm Re}\, k r_c \gtrsim 1000$, however, it is
appropriate to work with the correct TE formulas.)

%%%%%%%%%%%%%%%%%%%%%%%%%%%%%%%%%%%%%%%%%%%%%%%%%%%%%%%%%%%%%%%%%%%%%%%%%%%

%\section{Discussion}

In conclusion, we investigated 
the reflection coefficients 
at a curved refractive index boundary
by relating them to resonances in a circular 
dielectric disk
and derived analytic
expressions valid in the limit of large wavenumbers.
The deviations from Fresnel's laws can be explained
within geometric optics by incorporating
the Goos-H\"anchen effect.
In this work we concentrated on the wave field
confined by reflection inside the microcavity.
The Goos-H\"anchen shift also affects the
wave field outside the cavity \cite{herb,fresnel_derrest2,tran},
because the modified
angle of incidence results in
a change of the angle of refraction. 
It is desirable to investigate the implications on the
remarkable emission directionality
of non-circular devices \cite{jensnature,gmachlscience},
which sometimes departs substantially from
what is expected from geometric optics \cite{tureci}.

We gratefully acknowledge helpful discussions with
M.~Berry, S.~W.~Kim, J.~U.~N\"ockel, K.~Richter, R.~Schubert, and J.~Wiersig.

%%%%%%%%%%%%%%%%%%%%%%%%%%%%%%%%%%%%%%%%%%%%%%%%%%%%%%%%%%%%%%%%%%%%%%%%%%%

\end{document}